\documentclass[runningheads,a4paper]{llncs}

\usepackage{cmap}
\usepackage[T1]{fontenc}

\usepackage{graphicx}
\usepackage[ruled]{algorithm2e}
\usepackage{mathtools}
\usepackage{amssymb}
\SetKwRepeat{Do}{do}{while}%

\usepackage[ngerman,english]{babel}
\addto\extrasenglish{\languageshorthands{ngerman}\useshorthands{"}}

\usepackage[%
rm={oldstyle=false,proportional=true},%
sf={oldstyle=false,proportional=true},%
tt={oldstyle=false,proportional=true,variable=true},%
qt=false%
]{cfr-lm}
\usepackage[math]{blindtext}

\usepackage{cite}

\usepackage{paralist}

\usepackage{csquotes}

\usepackage{microtype}

\usepackage{url}
\makeatletter
\g@addto@macro{\UrlBreaks}{\UrlOrds}
\makeatother
\usepackage{xcolor}

\usepackage{pdfcomment}
\hypersetup{hidelinks,
   colorlinks=true,
   allcolors=black,
   pdfstartview=Fit,
   breaklinks=true}
\usepackage[all]{hypcap}

\usepackage[capitalise,nameinlink]{cleveref}
\crefname{section}{Sect.}{Sect.}
\Crefname{section}{Section}{Sections}

\usepackage{xspace}

\DeclareFontFamily{U}{MnSymbolC}{}
\DeclareSymbolFont{MnSyC}{U}{MnSymbolC}{m}{n}
\DeclareFontShape{U}{MnSymbolC}{m}{n}{
    <-6>  MnSymbolC5
   <6-7>  MnSymbolC6
   <7-8>  MnSymbolC7
   <8-9>  MnSymbolC8
   <9-10> MnSymbolC9
  <10-12> MnSymbolC10
  <12->   MnSymbolC12%
}{}
\DeclareMathSymbol{\powerset}{\mathord}{MnSyC}{180}

\begin{document}

\title{Anisotropic Radial Layout for Visualizing Centrality and Structure in Graphs }

\author{Mukund Raj \and Ross T. Whitaker}

\institute{University of Utah}

\maketitle

\begin{abstract}
This paper presents a novel method for layout of undirected graphs, where nodes (vertices) are
constrained to lie on a set of nested, simple, closed
curves.   Such a layout is useful to simultaneously display the structural
centrality and vertex distance information for graphs in many domains, including social networks. 
Closed curves are a more general constraint than the previously
proposed circles, and afford our method more flexibility to preserve vertex relationships compared to
existing radial layout  methods. The proposed approach 
modifies the multidimensional scaling (MDS) stress to include the estimation of a vertex depth or centrality field
as well as a term that
penalizes discord between structural centrality of vertices and their alignment with this carefully estimated field.   
We also propose a visualization strategy for the proposed layout and demonstrate its
effectiveness using three social network datasets.
\end{abstract}



\section{Introduction}\label{sec:intro}

\begin{figure}[!tb] \begin{center} \begin{tabular}{cc}
\includegraphics[ height = 1.75in]{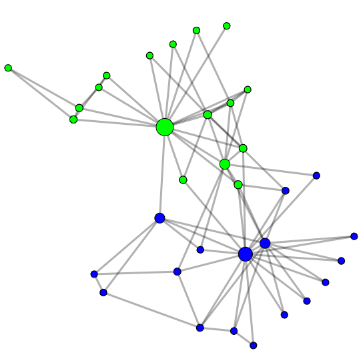} &
\includegraphics[ height = 1.75in]{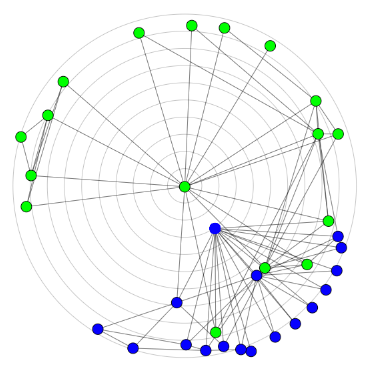} \\ (a)
& (b) \\ 
\multicolumn{2}{c}{\includegraphics[ height =
2.0in]{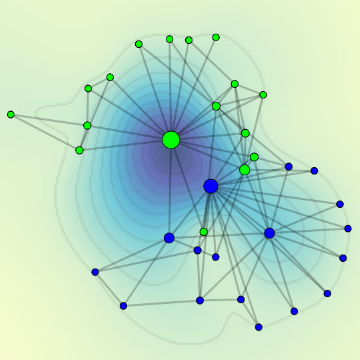}
\includegraphics[height = 1.5in]{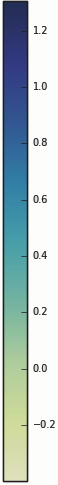}
} \\ \multicolumn{2}{c}{(c)} \\ \end{tabular}
\vspace{-12pt} \end{center} 
\caption{\label{fig:karate} Visualization of Zachary's karate club 
social network using (a) MDS, (b) Radial layout, and (c) Anisotropic radial layout. Node sizes 
encode betweenness centrality.} 
\end{figure}

Graphs are an important data structure that are used to represent relationships
between  entities in a wide range of domains.  An interesting aspect in graph
analysis  is the notion of (structural) \emph{centrality},  which pertains to
quantifying importance of entities (or vertices, nodes) within the context of the graph
structure as defined by it's relationships (or edges). The need to compute
centrality and convey it through visualization is seen in many areas, for
example,  in biology~\cite{schreiber2009generic},
transportation~\cite{brandes2009more} and  social
sciences~\cite{brandes2003communicating}. In this work, we propose a
method to visualize node centrality information in the context of overall graph
structure, which we capture through intervertex (graph theoretical) distances. 
The proposed method  determines a \emph{layout} (positions of nodes
on a 2D drawing) that meet the following two, often competing, criteria:
\begin{itemize}
\item \emph{Preservation of distances:} The Euclidean (geometrical) distances in the layout
should approximate, to the extent possible,  the graph theoretical distances between
the respective nodes.
\item \emph{Anisotropic radial monotonicity:} Along any ray traveling away from the position of the most central
node, nodes with a lower centrality should be placed geometrically further along the ray.
\end{itemize}
We also introduce a visualization strategy for the proposed layout that
further highlights the centrality and structure in the graph by using additional 
encoding channels, and demonstrate the benefits of our approach with real datasets (see~\cref{fig:karate} as an example).

Visualization methods for gaining insights from graph structured data are an
important and active area of research.  Significant efforts in this area are targeted toward
developing effective layouts. Layout methods can have various goals that
range from trying to reduce clutter and edge
crossings~\cite{brandes2004analysis} to faithfully representing the structure by
preserving the distances between nodes and topological
features~\cite{gansner2004graph}.  As positions are the best way to graphically
convey numbers~\cite{cleveland1984graphical},  layouts are also used to convey
numerically encoded measures of hierarchy or importance associated with 
nodes~\cite{dwyer2006ipsep,brandes2003communicating}.

Radial layouts have been shown to be an effective method to visually convey the relative {\em importance}
of nodes, where importance may be defined, for instance, by a node's
{\em centrality}~\cite{brandes2003communicating}.  The centrality of a node is a quantification of
 its importance in a graph by considering its various structural properties, 
 such as, connectedness, 
closeness to others, and role as an intermediary~\cite{freeman1978centrality,zweig2014network}.  
In conventional radial layouts, the distance
of nodes from the geometric center (origin) of the layout depends \emph{only}  on the
node's centrality, and nodes with a higher centrality value are placed closer
to the origin in the layout, often times forming rings or concentric circles.

Given a graph and centrality values associated with its nodes, several
approaches have  been proposed to determine a radial layout.  One line of work,
which deals with discrete centrality values, attempts to minimize edge
crossings~\cite{bachmaier2007radial}. Another approach, which also tackles
continuous centrality  values, involves optimizing a \emph{stress} energy
(\cref{subsec:mds}) by including a penalty for representation error (of graph distances) 
as well as deviation from radial
constraints~\cite{brandes2003communicating,brandes2009more}.  
The penalty acts a \emph{soft} constraint wherein the solution is allowed to deviate
from the constraint at the expense of increased local stress.  The literature shows that 
radial constraints may also be included as a \emph{hard}  constraint  by only
allowing  those solutions that satisfy the
constraints~\cite{dwyer2009constrained,baingana2014embedding,dwyer2009scalable}.

While state-of-the-art methods for radial graph layout do effectively convey node centrality,  the 
associate circular centrality constraints make it difficult to preserve other important, structural graph
characteristics such as distances, which, in turn, makes it difficult to
preserve the holistic structure of the graph.  On the other hand, despite being
effective in preserving the overall structure, general layout methods
such as multidimensional scaling are often fail to readily convey centrality
(e.g. by failing to ensure that structurally central nodes in the graph-theoretical sense appear
near the center of the layout and vice versa). In this
manuscript, we propose a method that simultaneously tackles both the above
issues.

The underlying idea for the proposed layout algorithm is that we can  relax the
constraint that requires nodes with similar centrality to lie on  a circle, and
instead, allow for such nodes to be constrained by a more general shape: a
simple closed  curve or \emph{centrality contour}.  Centrality contours are
nested isolevel curves on a smooth, radially decreasing estimate of  node
centrality values over a 2D field. We demonstrate that the
additional flexibility in placing the nodes afforded by the centrality
contours over circles, in conjunction with some additional visual cues in the
background, lets us achieve a better trade off than existing methods in conveying
centrality and general structure together.


\section{Background}
\label{sec:background}
In this section, we describe the various underlying technicalities that 
are relevant to the proposed method, and begin with some notation/definitions.

We define a weighted, undirected graph $G(V,E,W)$ as a set of vertices (or
nodes) $V$, a set of edges $E \subseteq V \times V$ and a set of edge weights,
$W:E \mapsto \mathbb{R}^+$, assigned to each edge. 
We define $n$ to be cardinality of node set; i.e., $n=|V|$. 
The graph-theoretical  distance (shortest-path along edges) between  two nodes $u$ and $v$ is denoted by $d_{uv}$. 
We denote a general position in a 2D layout as
${\bar{x}=(x,y)}$ and the Euclidean distance between two nodes $u$ and $v$ as
${\delta(\bar{x}_i,\bar{x}_j) =|| \bar{x}_u - \bar{y}_v ||_2}$.

\subsection{Centrality and Depth}
\label{subsec:depth}

The need to measure, and quantify, the importance of individual entities
within the context of a group occurs in many domains.  In graph
analytics, this need is addressed by \emph{centrality} indices, which are typically real-valued 
functions over the nodes of a graph~\cite{zweig2014network}. The specific properties that
qualify the  importance of nodes may depend on the application or data type, and 
several methods to compute centrality have been
proposed, such as degree centrality~\cite{freeman1978centrality}, 
closeness centrality~\cite{sabidussi1966centrality}, and betweenness
centrality~\cite{freeman1978centrality}.  While the emphasis of the various
centrality definitions can be different, they all share a common characteristic of depending
only on the \emph{structure} of the graph rather than parameters associated
with the nodes~\cite{zweig2014network}. For the examples in this paper we use
betweenness centrality due to its relevance to the datasets
(\cref{sec:results}). 

The {\em betweenness centrality} of a node, $v\in G$, is defined as
the percentage (or number) of shortest paths in the entire graph $G$ that pass through the node $v$.
As shown in work of Raj et al.~\cite{raj2017path}, barring instances of multiple geodesics, betweenness centrality is a special  case
of a more general notion of \emph{vertex depth} on graphs---a generalization of data depth to vertices on graphs. 
Data depth is a family
of methods from descriptive statistics  that attempts to quantify the idea of
centrality for ensemble data without any assumption  of the underlying
distribution.   Data depth methods often rely on the formation of
\emph{bands} from convex sets and the probability of a point lying within a randomly chosen band.  
The extension of band depth to graphs~\cite{raj2017path} relies on the convex closure of a set of points (via shortest paths), and thereby 
generalizes betweenness centrality by considering bands
formed by sets of nodes,  rather than only the shortest paths between pairs of nodes, and 
allows for a nonuniform probability distribution over the nodes of the graph.

\begin{figure}[!tb] \begin{center} \begin{tabular}{cc}
\includegraphics[height = 1.5in]{images/colorbar_lesmis}
\includegraphics[height = 1.75in]{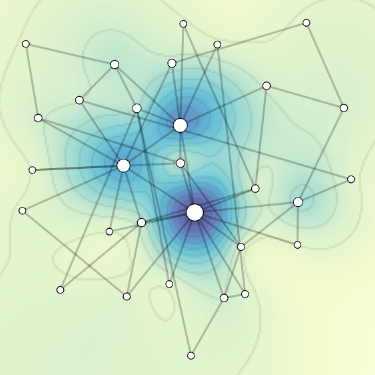} &
\includegraphics[height = 1.75in]{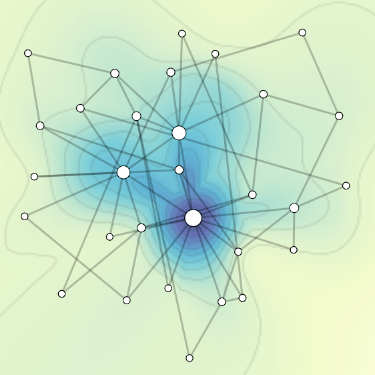} 
\includegraphics[height = 1.5in]{images/colorbar_lesmis}\\ (a)
& (b) \\ 

\end{tabular}
\vspace{-12pt} \end{center} 
\caption{\label{fig:background} An (a) \emph{interpolation field} for node centrality 
values, and (b) the associated (radially) \emph{monotonic field} for a
 30 node random graph generated using the Barabasi-Albert model. 
Node positions are determined using MDS and node sizes
encode betweenness centrality.} 
\end{figure}

In addition to graphs, data depth methods have been proposed for several  other
data types such as points in Euclidean space~\cite{tukey1975mathematics},
functions~\cite{lopez2009concept}, and
curves~\cite{pintado2014simplical,mirzargar2014curve}. Despite their  distinct
formulations, data depth methods are expected to share a  few common desirable
properties~\cite{Zuo00} such as: \begin{inparaenum}  \item maximum at geometric
center \item zero at infinity \item radial monotonicity \end{inparaenum}; which
make data depth an attractive basis for ensemble visualization
methods~\cite{rousseeuw1999bagplot,sun2011functional,mirzargar2014curve}. 
Graph centrality is a type of data depth on the nodes of a graph, and here we pursue layout methods 
that convey these depth properties.  

\subsection{Stress and Multidimensional Scaling (MDS)}
\label{subsec:mds}

Our proposed method is based
on a modification to the MDS objective function, and therefore we give a brief summary of MDS.  
MDS is family of methods that help
visualize the similarity  (or dissimilarity) between members in a data
set~\cite{borg2005modern}. Over the years,   MDS has been the foundation  for a
range of graph drawing algorithms that aim to achieve an isometry between graph
theoritical- and Euclidian distances between
nodes~\cite{kamada1989algorithm,brandes2009more}. From among various types of
MDS  methods that exist, here we consider \emph{metric MDS with distance
scaling}, which is popular in the graph drawing literature~\cite{gansner2004graph}
(see~\cref{fig:background}  for an example).


In the context of graph drawing, given a distance matrix based on graph-theoretical  
distance, the goal is to find node
positions ${X = \{\bar{x}_i:1 \leq i \leq n\}}$ that minimize
the following sum of squared residuals---also known as \emph{stress}: 
\begin{equation} \label{eqn:mds} \sigma(X) = \sum_{u,v}
w_{uv}\big( d_{uv} - ||\bar{x}_u - \bar{x}_v||_2 \big)^2 ,
\end{equation}
where $w_{uv} \geq 0$ is the weighting term for residual associated with pair
$u,v$. In the proposed work we employ a standard weighting scheme for graphs,
known as \emph{elastic scaling}~\cite{mcgee1966multidimensional}, by setting
$w_{uv}=d^{-2}_{uv}$. This gives preference to local distances by minimizing
\emph{relative} error rather than \emph{absolute} error during the optimization.


Node positions that minimize the objective (\cref{eqn:mds}) have been shown to
be visually pleasing and convey general structure of the graph~\cite{kamada1989algorithm}. 
Although, the state-of-the-art approach for optimizing the objective function 
is \emph{stress majorization}~\cite{gansner2004graph}, we 
employ standard \emph{gradient descent} because of its compatibility 
with the proposed modification to the objective (\cref{sec:method}). The gradient 
of the standard MDS objective is as follows~\cite{borg2005modern}:
\begin{equation}
\nabla \sigma(X) = 2VX - B(X)X 
\end{equation}
where matrices $V=(v_{ij})$ and $B=(b_{ij})$, with $1 \leq i,j \leq n$,
can be compactly represented as:
\begin{align*}
v_{ij} &= \begin{cases}
               -w_{ij} &\textrm{for }i \neq j\\
               \sum_{j=1,j\neq i}^n w_{ij}&\textrm{for }i = j
            \end{cases} 
            & 
 b_{ij} &= \begin{cases}
               -\frac{w_{ij}d_{ij}}{\delta(\bar{x}_i,\bar{x}_j)} &\textrm{for }i \neq j \textrm{ and } \delta(\bar{x}_i,\bar{x}_j) \neq 0\\
               0 &\textrm{for }i \neq j \textrm{ and }  \delta(\bar{x}_i,\bar{x}_j)=0\\
            \end{cases} \\
            & &
 b_{ii} &= -\sum_{j=1,j\neq i}^n b_{ij}.
\end{align*}

\subsection{Strictly Monotone and Smooth Regression}
\label{subsec:mono}

The proposed method also relies on the construction of a smooth and radially
decreasing approximation  of centrality values over a 2D field, which we call
the \emph{monotonic field} (\cref{fig:background}). The first part of this
construction  is an interpolation of centrality values of sparsely located nodes
on the layout to obtain a dense 2D field, which we call the \emph{interpolation
field} (\cref{fig:background}a). For this we use \emph{thin plate
splines}~\cite{bookstein1989principal} interpolation, a standard
technique for  interpolating unstructured data which produces optimally smooth fields.

The next part is to construct a radially monotonic approximation of the
interpolation field.   We devote the rest of this section to a brief description
of the method that we use for constructing this approximation (monotonic field), which is adapted from 
Dette et al~\cite{dette2006simple,dette2006strictly}. 

For a 1D function \cite{dette2006simple}, $m(t):[0,1] \rightarrow \mathbb{R}$, an elegant algorithm for computing its monotonic  
approximation $\hat{m}_A(t)$ proceeds as follows in \emph{two} steps~\cite{dette2006simple}:
\begin{itemize}
   \item \emph{Step 1 (Monotonization)}: Construct a density estimate from  sampled
values of input function $m$ and use it as input to compute  an
estimate of the inverse of the regression function $\hat{m}_A^{-1}$. 
\begin{equation}
\label{egn:mono}
\hat{m}_A^{-1}(t) = \frac{1}{Q \omega} \sum_{i=1}^Q \int_{\infty}^t K \Bigg( \frac{m\big(\frac{i}{Q}\big)-u}{\omega} \Bigg) du ,
\end{equation}
where $Q$ is the parameter controlling the sampling density, $K$ is a continuously differentiable and symmetric 
kernel, and $\omega$ is the bandwidth. Here, $\hat{m}_A^{-1}$ is a strictly \emph{increasing} 
estimate of $m^{-1}$, however, we can easily obtain a strictly \emph{decreasing} estimate 
by reversing the limits on the integral in~\cref{egn:mono}.
\item \emph{Step 2 (Inversion)}: Obtain the final estimate of $\hat{m}_A$ by 
numerically inverting $\hat{m}_A^{-1}$.
\end{itemize}


In order to obtain an approximation to a 2D function that is monotonic along
radial lines emanating from the deepest or most central node, we use a polar
coordinate representation of the field. We build the polar representation  by
sampling the interpolation field along 360 evenly spaced, center outward rays.
The idea is to repeatedly monotonize the interpolation field with respect to a
single variable i.e., for a fixed value of the angular coordinate, obtain a (1D)
estimate that is strictly decreasing along the radial coordinate. We then repeat
this process, successively monotonizing 1D functions that correspond to each
value of angular coordinate in its (discrete) domain; see~\cref{fig:background}b
for an example of the resulting monotonic field. The spline
interpolation is smooth, and by the properties of the monotonic approximation
(see~\cite{dette2006strictly}), the resulting monotonic field is smooth (except
at origin, where polar the coordinates maybe nonsmooth).

\section{Method}
\label{sec:method}

\begin{figure}[!tb] \begin{center} \begin{tabular}{cc}
\includegraphics[height = 1.5in]{images/colorbar_lesmis}
\includegraphics[height = 1.75in]{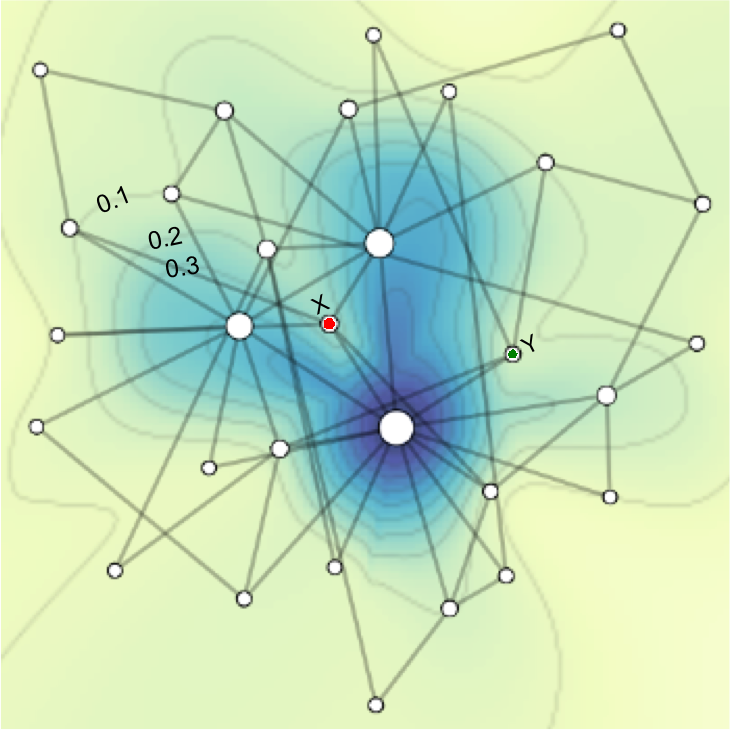} &
\includegraphics[height = 1.75in]{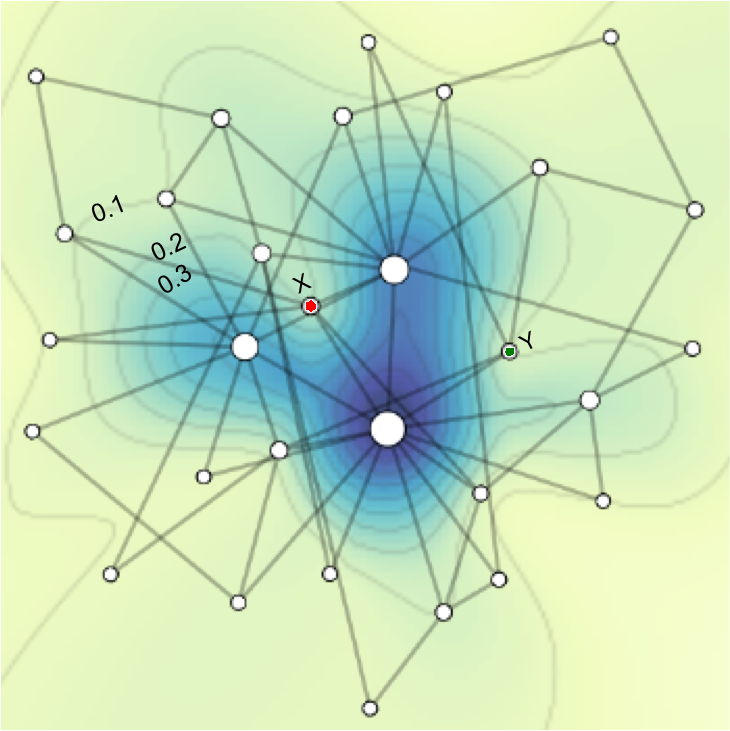} 
\includegraphics[height = 1.5in]{images/colorbar_lesmis}
\\ (a)
& (b) \\ 
\includegraphics[height = 1.5in]{images/colorbar_lesmis}
\includegraphics[height = 1.75in]{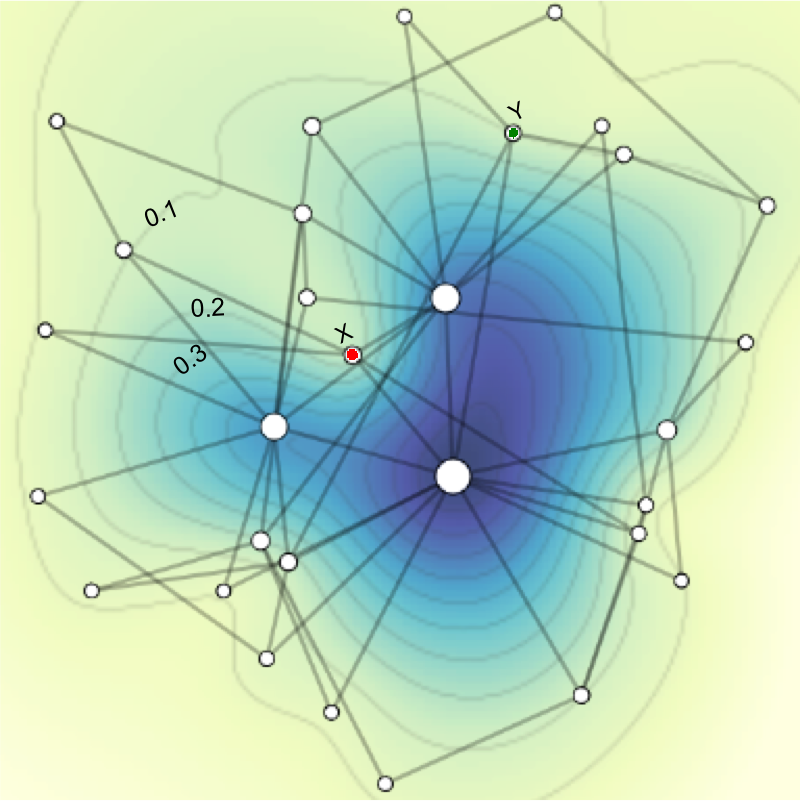} &
\includegraphics[height = 1.75in]{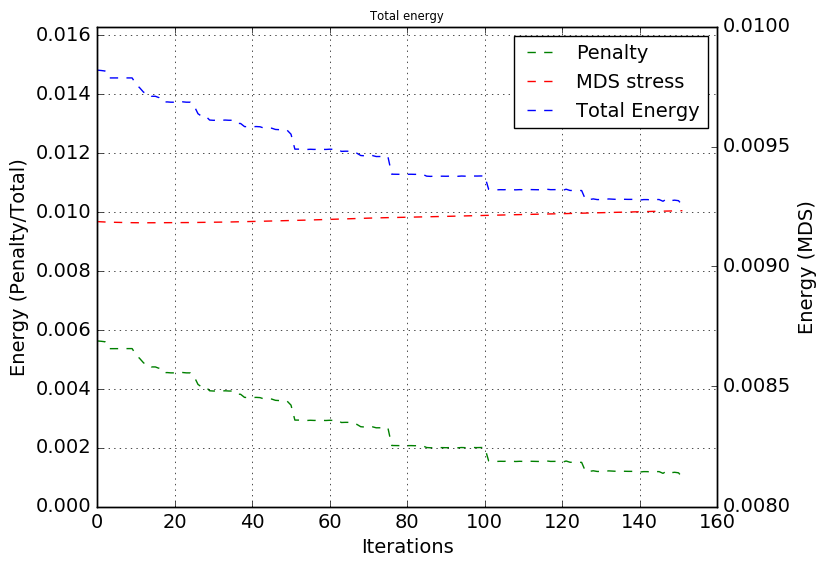} \\ (c)
& (d) \\ 
\end{tabular}
\vspace{-12pt} \end{center} 
\caption{\label{fig:method} Sensitivity of anisotropic radial layout to penalty weights for the graph in~\cref{fig:background}: 
(a) $w_{\rho}=0.1$, (b) $w_{\rho}=1$, (c) $w_{\rho}=10$; centrality contours with isovalues 0.1, 0.2 and 0.3 
as well as nodes \emph{X} (red) and \emph{Y} (green) with centrality values 0.2 and 0.1 are identified,
and (d) A typical plot of objective energy during the optimization 
process ($w_{\rho}=1$).}
\end{figure}

Here we describe our method in two parts.  First is the layout algorithm
(\cref{subsec:arl}), and second is a visualization strategy (\cref{subsec:vis})
that complements the layout to simultaneously convey graph structure and node centrality.

\subsection{Anisotropic Radial Layout}
\label{subsec:arl}

In addition to preserving the graph-theoretical distances, we also aim to place
every node on a radially monotonic approximation of a centrality field---called
the \emph{monotonic field} (\cref{subsec:mono})---such that the value of the
field at the location of the node is equal to the centrality value of the node.
We accomplish this by modifying the (distance preserving) MDS objective or \emph{stress} 
(\cref{subsec:mds}) to incorporate the
following penalty term, which penalizes the deviation of monotonic field values
from the node centrality values:

\begin{equation}
\rho(X) = \big( M_{X,\bar{c}}(X) - \bar{c}\big)^2
\end{equation}
where $\bar{c} \in \mathbb{R}^n$ is a vector of node centrality values and ${X \in
\mathbb{R}^{n  \times 2} =\{\bar{x}_i:1 \leq i \leq n\}}$ denotes associated node positions.
$M_{X,\bar{c}}(X) \in \mathbb{R}^n$ denotes a vector of values of the 2D
monotonic field at locations $X$. 
The symbols in the subscript ($X$ and $\bar{c}$) denote the use of node positions and centrality values 
in the construction of the monotonic field.  In the limiting case where where the
\emph{interpolation field} (\cref{subsec:mono}) itself is monotonic,  the value of this
penalty term drops to zero. Our final objective is a sum of the MDS stress 
and the above penalty term, and can be stated as follows:
\begin{equation}
\label{eqn:arl}
\gamma \big(X\big) = \underbrace{\sigma(X)}_{\textrm{MDS stress}} +  \quad w_{\rho} \; \rho(X)
\end{equation}
where $w_{\rho}$ is a weighting factor that controls the influence of the penalty, with 
respect to the MDS stress. 
The gradient of the modified objective above is obtained as:
\begin{equation}
\label{eqn:arlgrad}
\nabla \gamma \big(X \big) = \nabla \sigma (X) + w_{\rho} \times \underbrace{2 \big( M_{X,\bar{c}}(X) - \bar{c}\big) \odot \nabla M_{X,\bar{c}}(X)}_{\nabla \rho(X)},
\end{equation}
where $\odot$ denotes element wise product.   It is difficult to compute the
gradient of $ M_{X,\bar{c}}(X)$  because of dependence of  $M$ on $X$ 
and the associated process for monotonic approximation. 
Therefore, we let the field lag, and treat
$X$ (in subscript) as a constant when numerically approximating the  gradient of
$M$. We deal with the resulting accumulation of error by recomputing the depth
field after a fixed number of iterations, or \emph{lag}, denoted by $\ell$.

The parameters $w_{\rho}$ and $\ell$ need to be chosen carefully.
$w_{\rho}$ needs to be set to find a balance between preserving the intrinsic
graph structure and ensuring that the centrality of nodes match the field
value at their position. \cref{fig:method}a-c show, respectively,
results of a \emph{small} $w_{\rho}$ unable to move nodes to
appropriate positions with regard to the field (observe nodes \emph{X},\emph{Y}), an \emph{intermediate} $w_{\rho}$, and a
\emph{large} $w_{\rho}$ resulting in unnecessary structural distortion with regard to 
initial positions (observe node \emph{Y}).
The parameter $\ell$
controls the lag of the monotonic field; if $\ell$ is too small, the frequent updates can lead to instabilities, 
while values that are too large can cause slow convergence. A
typical energy profile during optimization is shown in \cref{fig:method}d; where
the sharp changes in the total energy correspond to the updates of the monotonic
field. We  encourage the layout to be as similar as possible to the MDS
layout by  initializing the node positions as determined by an \emph{unmodified}
MDS objective~\cite{gansner2004graph}.
The entire process, as summarized in \cref{fig:arlalgo}, iterates until updates no longer result in significant changes to node positions.

The computational complexity of a single iteration is $\mathcal{O}(n^3)$ due to
the step of computing the monotonic field which involves interpolation using thin
plate spline.  However, we only update the field once every $\ell$
iterations. This leads to a complexity of $\mathcal{O}(n^2)$ (same as MDS) for
a large majority of iterations.

\begin{algorithm}[!t]
\SetAlgoLined
 \KwIn{Graph $G=\{V,E,W\}$, maximum number of iterations $k \in \mathbb{N}$, depth field lag $\ell$, step size $\alpha$, weighing factor $w_{\rho}$}
 \KwOut{Positions ${X = \{\bar{x}_i:1 \leq i \leq n\}}$ for all $v_i \in V$}
$n \leftarrow \vert V \vert$\\
 $X_0 \leftarrow$ initialize node positions using MDS  \tcc*{(\cref{subsec:mds})}
 $\bar{c} \in \mathbb{R}^n \leftarrow$ compute graph centrality values for $v_i \in V$ \\
$j \leftarrow -1$ \tcc*{index to keep track of field updates}
 \For{ $t=1,\ldots, k$ }{
	\If{$t \mod \ell=0$ }{
 		$j \leftarrow j+1$\\
 		$X_j \leftarrow X_t$ \\
     $M_{X_j,\bar{c}}(X_t) \leftarrow $ compute monotonic field \tcc*{(\cref{subsec:mono})}
      }
     $X_{t+1} \leftarrow X_{t} - \alpha \Big( \nabla \sigma (X_{t}) + w_{\rho} \times 2 \big( M_{X_j,\bar{c}}(X_t) - \bar{c}\big) \odot \nabla M_{X_j,\bar{c}}(X_t) \Big) $\tcc*{gradient update step (\cref{subsec:arl}) }
   }
\caption{\label{fig:arlalgo}Layout with anisotropic radial constraints}
\end{algorithm}

\subsection{Visualization}
\label{subsec:vis}

\begin{figure}[!t] \begin{center} \begin{tabular}{cc}
\includegraphics[ height = 1.75in]{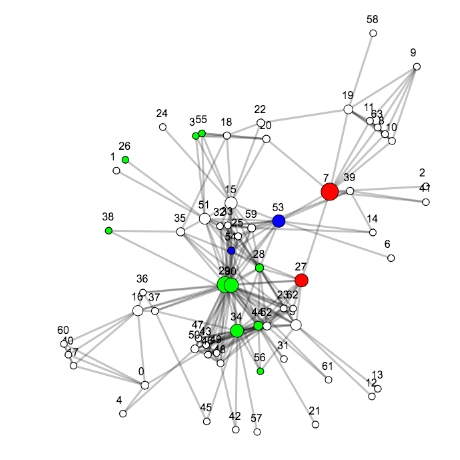} &
\includegraphics[height = 1.75in]{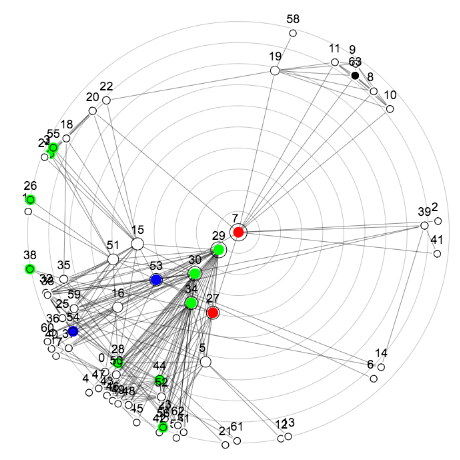} \\ (a)
& (b) \\ 
\multicolumn{2}{c}{\includegraphics[ height =
2in]{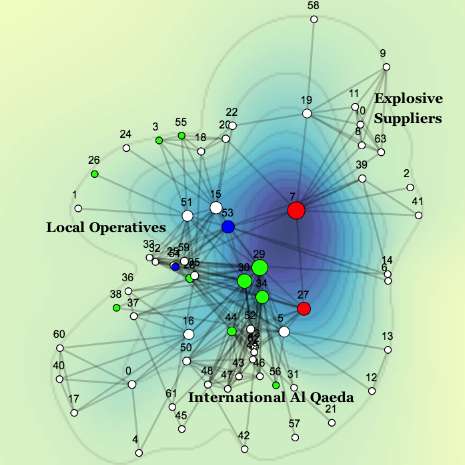}
\includegraphics[height = 1.5in]{images/colorbar_lesmis}
} \\ \multicolumn{2}{c}{(c)} \\
 \end{tabular}
\vspace{-12pt} \end{center} 
\caption{\label{fig:terror} Network of terrorists and affiliates connected to the 2004 Madrid train bombing using (a) MDS, (b) radial layout, (c) anisotropic radial layout.} 
\end{figure}

In this layout, nodes are constrained to lie on level sets of centraility, which are \emph{general} closed curves, 
rather than circles, and the shapes of these curves depend on the structure of the graph.  Therefore, we can 
help interpretability of the layout and reduce
cognitive load for the user by providing additional cues for shapes of these curves. 
We provide cues in the form of faded
renderings of centrality contours (isolines on the monotonic field) and a
monotonic field colormap in the background.   
The radial monotonicity described in~\cref{subsec:arl} ensures
that the  contours are nested curves that enclose a \emph{common} maxima (at origin);
leading to a bijective
mapping between contours and centrality values, and pushing nodes to lie on the \emph{unique} contour that corresponds to
their centrality.  In this paper, we normalize node
centrality to fall between 0 and 1; and show 10 contour curves that evenly span
this range.  We also  use node size as an extra encoding channel for centrality
---in addition to location---to further highlight the centrality structure.
We can, of course, use the size channel to encode centrality even with the
standard MDS layout, however, that approach can lead to the issue of
conflicting centrality cues from  size and location channels
(see image (a) in \cref{fig:karate,fig:terror,fig:lesmes}).

\section{Results}\label{sec:results}
\subsection{Zachary's karate club}
The Zachary's karate club graph is a well known data set that is a social network of
friendships in a karate club at a US university, as recorded during a
study~\cite{zachary1977information}. This graph  contains 34 nodes, each
representing an individual, and 78 unweighted edges that represent  a friendship
between the associated individuals (\cref{fig:karate}). During the period of
observation  , a conflict between two key members, identified as the
``administrator'' and  ``instructor'', leads to a split in the club, giving  it
an interesting two cluster structure. In \cref{fig:karate}, nodes representing
members who are part of the instructor's and administrator's groups are  drawn
in green and blue, respectively. 

\cref{fig:karate} shows three different visualizations of the karate club network:
MDS,  radial layout (from \cite{brandes2009more}),  and anisotropic radial layout (ARL). We can make a few
observations from the visualizations.  While MDS does a good
job of preserving the two clusters, it is does not unambiguously convey
centrality. On the other hand,  radial layout clearly showcases the centrality
at the expense of dispersing the clusters by distorting distances among 
their nodes, thereby obscuring their internal structure.
 We see that ARL is able to largely preserve the structure seen in
MDS with clearly  distinguishable clusters, and also clearly convey the
centrality information.  While radial layout pushes the instructor's group far
away due to low betweenness centrality,   ARL lets them remain close by
\emph{bringing in} the outermost contour toward to the group instead.   Similarly,
the administrator is also allowed to remain closer to their group by the
protrusion of the inner contours, which enclose the most central nodes, toward
the  administrator.


\subsection{Terrorist network from 2004 Madrid train bombing}
\begin{figure}[!tb] \begin{center} \begin{tabular}{cc}
\includegraphics[ height = 1.75in]{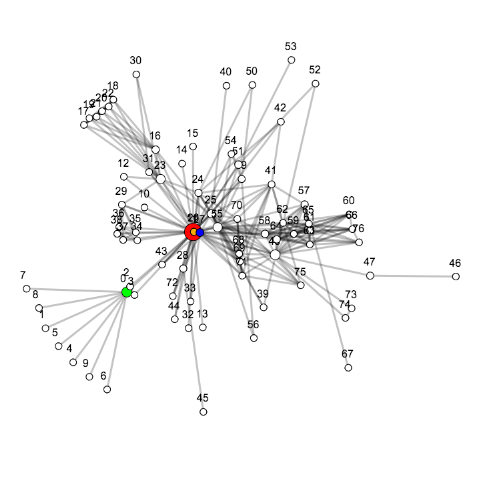} &
\includegraphics[ height = 1.75in]{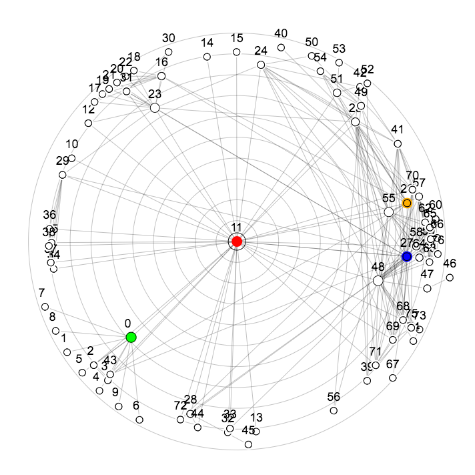} \\ (a)
& (b) \\ 
\multicolumn{2}{c}{
\includegraphics[height = 2.0in]{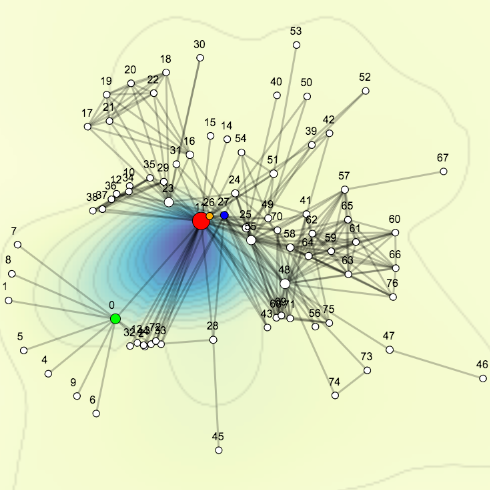}
\includegraphics[height = 1.5in]{images/colorbar_lesmis}
} \\ \multicolumn{2}{c}{(c)} \\ \end{tabular}
\vspace{-12pt} \end{center} 
\caption{\label{fig:lesmes} Coappearance network for characters in the novel Les Miserables using (a) MDS, (b) radial layout,  (c) anisotropic radial layout.} 
\end{figure}

\cref{fig:terror} shows visualizations of a network of individuals connected  to
the bombing of trains in Madrid on March 11, 2004. This data was originally
compiled by Rodriguez~\cite{rodriguez2005march} from newspaper  articles that
reported on the subsequent police investigation. There are 64 nodes that
represent suspects and their relatives, and 243 edges that have  weights ranging
from 1 to 4 which represent  an aggregated  strength of connection based on
various parameters such as contact, kinship, ties to Al Qaeda,
etc~\cite{hayes2006connecting}. In \cref{fig:terror}, (as well as
\cref{fig:lesmes}), distances between nodes are  related inversely to edge
weights. In the visualization, we identify nodes using numbers to avoid text clutter,
however, we include a mapping to names of individuals represented by the nodes 
in the Appendix.     


Rodriguez~\cite{rodriguez2005march} identifies several key suspects as follows:
ring leaders (marked in blue in \cref{fig:terror}), members of a  field
operating group who were closely involved with the actual carrying out the
attack (green),  intermediaries (red), as well as suspects with local roots,
ties to  foreign Al Queda, and those who supplied explosives. On comparing the
visualizations  in \cref{fig:terror} we see that ARL (\cref{fig:terror}c) is able to better preserve
the structure and cohesiveness of the core members of the field operating group
in comparison to the radial layout (\cref{fig:terror}b).   
 Critically, a key mastermind in this event, despite having
a low centrality (due to communicating often through an
intermediary), is allowed to be close to the center in the ARL.
This arrangement, possible due to the ability of centrality contours to adapt to the circumstance, 
preserves the close association between the masterminds
that is lost in the radial layout.     We also see  that the
flexibility of contours in ARL preserves the locality of various groups, which allows
us to  see the role of intermediaries with high centrality in acting as a bridge
between various groups.

\subsection{Coappearance network for characters in Les Miserables}
The third dataset is a graph of character associations in the famous French
novel  Les Miserables (\cref{fig:lesmes})\cite{knuth1993stanford}. This graph
consists of 77 nodes, each representing  a character in the novel, and 254
weighted edges where the weights represent the  number of chapters
that feature both characters associated with an edge.  



We see the that the main protagonist {\em Valjean} (marked in red) is placed
prominently in all three visualizations (\cref{fig:lesmes}). However, other
key characters in the plot such as {\em Inspector Javert} (blue)
and {\em Cosett} (orange), who do not appear often with characters other than the
protagonist (and thus have low betweenness centrality) are treated differently.
While the radial layout relegates them to the periphery (far from Valjean) (\cref{fig:lesmes}b), 
MDS (\cref{fig:lesmes}a) paints a conflicting
picture with regard to their centrality, e.g.,  Cosett's node almost overlaps
with Valjean despite its low centrality. In contrast, the proposed ARL (\cref{fig:lesmes}c)
is able to coherently convey the low centrality  of the Inspector Javert and
Cosett, as well as, their closeness to Valjean. The above issue of distance
distortion appears to be a frequent occurrence in the radial layout due to many
characters  who have a low centrality value causing them to end up being packed
in the outer  periphery. A case of contrast is that of the character {\em Bishop
Myriel} (green) who despite being associated with several characters, is only seen with Valjean
once.


\section{Discussion}\label{sec:discussion}

This paper describes an energy-based layout algorithm for graphs, called {\em
anisotropic radial layout}, which conveys structural centrality using
\emph{anisotropic}, radial constraints, while also preserving approximate
distances (or structure) in the graph. In contrast to existing methods for
conveying node centrality which employ an \emph{isotropic} centrality field~\cite{brandes2009more,baingana2014embedding}, the
proposed method determines an \emph{anisotropic} centrality field on which to project
nodes. While the energy minimization strategy described in this paper allows the
solution  to deviate from constraints, one can enforce hard constraints by
adding a post processing step that projects nodes onto the closest position 
on their associated isocontour.

The key implication of the anisotropic centrality field in our method is that more
central nodes are allowed to be placed further from origin than less central
nodes---without an energy penalty---if they do not lie on a common ray; which
aids our objective of achieving a better balance between visual representations
of centrality and structure than possible with existing methods. Our objective
differs from other prior work that use centrality or continuous fields to visualize
structure of  dense graphs~\cite{van2008centrality,van2003graphsplatting}.

\section{Appendix}\label{sec:appendix}

\subsection{Name associations for nodes in the terrorist network (\cref{fig:terror}) }

\begin{minipage}{2.5in}
\begin{tabular}{cc}
Node ID & Name\\
\hline
0 & Jamal Zougam  \\ 
1 & Mohamed Bekkali  \\ 
2 & Mohamed Chaoui \\ 
3 & Vinay Kholy \\ 
4 & Suresh Kumar  \\ 
5 & Mohamed Chedadi  \\ 
6 & Imad Eddin Barakat \\ 
7 & Abdelai Benyaich \\ 
8 & Abu Abderrahame \\ 
9 & Omar Dhegayes  \\ 
10 & Amer Aii \\ 
11 & Abu Musad Alsakaoui \\ 
12 & Mohamed Atta \\ 
13 & Rami Binalshibh \\ 
14 & Mohamed Belfatmi \\ 
15 & Said Bahaji \\ 
16 & Al Amrous \\ 
17 & Galeb Kalaje \\ 
18 & Abderrahim Zbakh \\ 
19 & Farid Oulad Ali \\ 
20 & Jos Emilio Sure  \\ 
21 & Khalid Ouled Akcha  \\ 
22 & Rafa Zuher \\ 
23 & Naima Oulad Akcha \\ 
24 & Abdelkarim el Mejjati \\ 
25 & Abdelhalak Bentasser \\ 
26 & Anwar Adnan Ahmad \\ 
27 & Basel Ghayoun \\ 
28 & Faisal Alluch \\ 
29 & S B Abdelmajid Fakhet \\ 
30 & Jamal Ahmidan \\ 
31 & Said Ahmidan \\ 
32 & Hamid Ahmidan \\ 
33 & Mustafa Ahmidan \\ 
34 & Antonio Toro  \\ 
 \end{tabular}
\end{minipage}
\begin{minipage}{2.5in}
 \begin{tabular}{cc}
Node ID & Name\\
\hline
35 & Mohamed Oulad Akcha \\ 
36 & Rachid Oulad Akcha \\ 
37 & Mamoun Darkaanli \\ 
38 & Fouad El Morabit Anghar \\ 
39 & Abdeluahid Berrak \\ 
40 & Said Berrak \\ 
41 & Waanid Altaraki Almasri \\ 
42 & Abddenabi Koujma \\ 
43 & Otman El Gnaut \\ 
44 & Abdelilah el Fouad \\ 
45 & Mohamad Bard Ddin Akkab  \\ 
46 & Abu Zubaidah \\ 
47 & Sanel Sjekirika  \\ 
48 & Parlindumgan Siregar \\ 
49 & El Hemir \\ 
50 & Anuar Asri Rifaat \\ 
51 & Rachid Adli  \\ 
52 & Ghasoub Al Albrash \\ 
53 & Said Chedadi \\ 
54 & Mohamed Bahaiah \\ 
55 & Taysir Alouny \\ 
56 & OM Othman Abu Qutada  \\ 
57 & Shakur \\ 
58 & Driss Chebli \\ 
59 & Abdul Fatal \\ 
60 & Mohamed El Egipcio \\ 
61 & Nasredine Boushoa \\ 
62 & Semaan Gaby Eid \\ 
63 & Emilio Llamo \\ 
64 & Ivan Granados  \\ 
65 & Raul Gonales Pere \\ 
66 & El Gitanillo \\ 
67 & Mouta Almallah \\ 
68 & Mohamed Almallah \\ 
69 & Yousef Hichman \\ 
 \end{tabular}    
\end{minipage}

\subsection{Name associations for nodes in Les Miserables network (\cref{fig:lesmes})}

\begin{minipage}{1.85in}
\begin{tabular}{cc}
Node ID & Name\\
\hline
0 & Myriel \\ 
1 & Napoleon \\ 
2 & MlleBaptistine \\ 
3 & MmeMagloire \\ 
4 & CountessDeLo \\ 
5 & Geborand \\ 
6 & Champtercier \\ 
7 & Cravatte \\ 
8 & Count \\ 
9 & OldMan \\ 
10 & Labarre \\ 
11 & Valjean \\ 
12 & Marguerite \\ 
13 & MmeDeR \\ 
14 & Isabeau \\ 
15 & Gervais \\ 
16 & Tholomyes \\ 
17 & Listolier \\ 
18 & Fameuil \\ 
19 & Blacheville \\ 
20 & Favourite \\ 
21 & Dahlia \\ 
22 & Zephine \\ 
23 & Fantine \\ 
24 & MmeThenardier \\ 
25 & Thenardier \\ 
 \end{tabular}
\end{minipage}
\begin{minipage}{1.85in}
\begin{tabular}{cc}
Node ID & Name\\
\hline
26 & Cosette \\ 
27 & Javert \\ 
28 & Fauchelevent \\ 
29 & Bamatabois \\ 
30 & Perpetue \\ 
31 & Simplice \\ 
32 & Scaufflaire \\ 
33 & Woman \\ 
34 & Judge \\ 
35 & Champmathieu \\ 
36 & Brevet \\ 
37 & Chenildieu \\ 
38 & Cochepaille \\ 
39 & Pontmercy \\ 
40 & Boulatruelle \\ 
41 & Eponine \\ 
42 & Anelma \\ 
43 & Woman \\ 
44 & MotherInnocent \\ 
45 & Gribier \\ 
46 & Jondrette \\ 
47 & MmeBurgon \\ 
48 & Gavroche \\ 
49 & Gillenormand \\ 
50 & Magnon \\ 
51 & MlleGillenormand \
 \end{tabular}
\end{minipage}
\begin{minipage}{1.85in}
\begin{tabular}{cc}
Node ID & Name\\
\hline
52 & MmePontmercy \\ 
53 & MlleVaubois \\ 
54 & LtGillenormand \\ 
55 & Marius \\ 
56 & BaronessT \\ 
57 & Mabeuf \\ 
58 & Enjolras \\ 
59 & Combeferre \\ 
60 & Prouvaire \\ 
61 & Feuilly \\ 
62 & Courfeyrac \\ 
63 & Bahorel \\ 
64 & Bossuet \\ 
65 & Joly \\ 
66 & Grantaire \\ 
67 & MotherPlutarch \\ 
68 & Gueulemer \\ 
69 & Babet \\ 
70 & Claquesous \\ 
71 & Montparnasse \\ 
72 & Toussaint \\ 
73 & Child \\ 
74 & Child \\ 
75 & Brujon \\ 
76 & MmeHucheloup \\ 
&  \\ 
 \end{tabular}
\end{minipage}

\paragraph{}
\noindent
\textbf{Acknowledgments.} This work was supported by National Science Foundation (NSF) grant IIS-1212806.

\bibliographystyle{splncs03}
\bibliography{paper}{}

\end{document}